\documentclass[superscriptaddress,twocolumn,prb]{revtex4}

\usepackage{graphicx}
\usepackage{subfigure}
\usepackage{amsmath}

\begin{document}


\title{Quantum manipulation via atomic-scale 
magnetoelectric effects}
\author{Anh T. Ngo}
\affiliation{Department of Physics, and 
         Nanoscale and Quantum Phenomena Institute, 
         Ohio University, 45701 USA}
\author{Javier Rodriguez-Laguna}
\affiliation{ICFO-Institut de Ciencies Fotoniques, 
             Barcelona, Spain}
\author{Sergio E. Ulloa}
\affiliation{Department of Physics, and 
         Nanoscale and Quantum Phenomena Institute, 
         Ohio University, 45701 USA}
\author{Eugene H. Kim}
\affiliation{Instituto de F\'isica T\'eorica, UAM-CSIC, 
	 Madrid 28049, Spain}
\affiliation{Department of Physics, University of Windsor,
         Windsor, Ontario, Canada N9B 3P4}

\begin{abstract}
Magnetoelectric effects at the atomic scale are 
demonstrated to afford unique functionality.  
This is shown explicitly for a quantum corral 
defined by a wall of magnetic atoms deposited on a 
metal surface where spin-orbit coupling is observable.  
We show these magnetoelectric effects allow one to control 
the properties of systems placed inside the corral as well 
as their electronic signatures;
they provide alternative tools for probing electronic
properties at the atomic scale.
\end{abstract}

\maketitle



It has long been appreciated that interesting and 
unique properties arise from the coupling between 
the charge and magnetic degrees of freedom in materials.   
Advances in our understanding of the physics regulating
these properties have given rise to systems with a 
number of practical applications.
Prominent examples include multiferroic materials, 
which exhibit simultaneous and cooperative ferroelectric 
and magnetic ordering,\cite{multiferroic} as well as 
giant\cite{gmr} and colossal\cite{cmr} magnetoresistance 
materials, which exhibit substantial changes in their 
electronic transport with the application of small magnetic 
fields.  
These and other {\sl magnetoelectric} (ME) {\sl effects}
--- coupling charge and magnetic degrees of freedom ---
provide unique functionalities and hold promise for novel 
device applications.\cite{magnetoapp}

With the continuing drive to miniaturize electronic 
devices, there is interest in better controlling or/and 
enhancing the functionality of nanoscale systems; there 
is particular interest in novel devices that utilize 
phenomena inherent/unique to these nanometer 
scales.\cite{nanodevices}
Here, we predict unique functionality in a nanoscale 
device arising from the coupling of charge and magnetic 
degrees of freedom in the ultimate miniaturization, 
namely where devices are built 
atom-by-atom.\cite{nanodevices,corralcomment}  
We demonstrate that the interplay of spin-orbit coupling 
and electronic scattering results in ME effects which 
enable exquisite control at the atomic scale.   
This control provides a powerful tool for manipulating 
the response of quantum systems, adding desirable 
functionalities not only for fundamental studies, but 
also for possible future devices built in a 
``bottom up" approach.



Our device system consists of a quantum corral (QC) 
on a metal surface with spin-orbit coupling (SOC) 
(e.g. Au(111) \cite{exptAu}), where the QC's wall 
is made of magnetic atoms.
We demonstrate the possibility of controlling the 
electronic properties of the QC by changing the 
magnetization of the atoms forming the QC's wall;
we show that these ME effects allow one to control
the properties of systems placed inside the QC as 
well as their electronic signatures.
This control provides powerful alternative tools for
probing and manipulating electronic properties at the
atomic scale.

The Hamiltonian for the system has the form 
$\hat{H}$=$\hat{H}_{\rm QC}$+$\hat{H}_1$, where
$\hat{H}_{\rm QC}$ describes the QC, and $\hat{H}_1$ 
describes a system we place inside the QC, whose 
properties will be controlled and/or probed (see below).
We describe the QC by the Hamiltonian
$\hat{H}_{\rm QC}$=$\hat{H}_0$+$\hat{V}$, 
where $\hat{H}_0$ describes a two-dimensional electron gas 
(2DEG) with SOC, and $\hat{V}$ is a scattering potential 
describing the QC's wall.
The Hamiltonian for the 2DEG is
\begin{equation}
 \hat{H}_0 = \frac{1}{2m^*} {\bf p}^2 + \lambda~ \hat{z} 
  \cdot \left( {\bf p} \times \overline{\sigma} \right) 
\label{2DEGham}
\end{equation}
where ${\bf p}$ is the momentum operator of the 2DEG, 
$\{ \sigma^{\mu} \}$ are the Pauli matrices, $m^*$ is
the electron's band mass, and $\lambda$ parameterizes 
the SOC.
As described above, we are interested in the case where 
the QC's wall is made of magnetic atoms --- 
being interested in the system's low-energy properties,
we treat the atoms as a collection of $s$-wave 
scatterers;\cite{scatteringbook}
we take
\begin{equation}
 \hat{V} = \sum_i \left( V_0 
 + \frac{J}{2} \overline{\tau}_i \cdot {\overline \sigma} \right)
    \delta({\bf r} - {\bf r}_i)  \ ,
\label{potential}
\end{equation}
where ${\bf r}$ is the position operator of the 2DEG,
$V_0$ describes the potential scattering, 
$J$ is the exchange coupling between the 
(magnetic) atoms and the 2DEG,
$\overline{\tau}_i$ is the spin operator of the 
i$^{th}$ atom,
and the atoms are located at the positions 
$\{ {\bf r}_i \}$. \cite{SI}
The form(s) of $H_1$ will be specified subsequently.



In this work, we will be interested in the case where 
the atoms of the QC's wall are ferromagnetically (FM) 
ordered --- we assume the atoms' moments are sufficiently 
large and treat them as classical variables:
$(J/2)\langle \overline{\tau}_i \rangle$=${\bf M}$.
The physical quantity of interest is the electronic 
local density of states (LDOS) in the QC, $A({\bf r},\omega)$.
[The differential conductance measured in scanning tunneling 
microscopy is proportional to $A({\bf r},\omega)$.\cite{mahan}]
This is obtained from the QC's retarded Green's function (GF)
$G({\bf r},{\bf r}'; \omega)$ via
\begin{equation}
 A({\bf r},\omega) = -\frac{1}{\pi}~ {\rm Im}\left\{ 
  {\rm Tr}[G({\bf r}, {\bf r}; \omega)] \right\} \ ,
\end{equation}
where\cite{mahan,hewson}
\begin{equation}
 G({\bf r},{\bf r}'; \omega) 
   = G_0({\bf r},{\bf r}'; \omega)
   + G_0({\bf r},{\bf R}_0; \omega) \hat{T}(\omega)
     G_0({\bf R}_0,{\bf r}'; \omega)  \ .
\label{fullgreens}
\end{equation}
In Eq.~\ref{fullgreens}, the $T$-matrix $\hat{T}(\omega)$ describes 
the influence of $H_1$.  Furthermore, $G_0({\bf r},{\bf r}'; \omega)$ 
is the bare GF of the QC, i.e. the GF in the absence of 
$\hat{H}_1$.\cite{SI}

In what follows, we choose $(\pi \rho_0) V_0$=0.3 and 
$(\pi \rho_0) |{\bf M}|$=0.5 ($\rho_0$=$m^*/2\pi$);\cite{SI} 
we consider physically reasonable values of the parameters for the 
2DEG:\cite{SOnanolett} a Fermi energy $E_F$=0.5eV, $m^*$=0.26$m_e$
($m_e$ is the bare electron mass), and 
$\lambda$=4$\times$$10^{-11}$eV$\cdot$m.
The results we show are for an elliptical QC with 40 atoms, similar 
to what has been realized experimentally:\cite{corralexp}
$(x/a)^2$+$(y/b)^2$=$R^2$ with $R$=57.22\AA, $a/b$=1.5,
and $(\pm c,0)$=$(\pm \sqrt{a^2 - b^2},0)$ being the ellipse's 
foci; 
we will also comment about results obtained for a circular QC.
It should be stressed the results we report are robust ---
changes in the relative size and phase of the ratio ${\bf M}/V_0$ 
have only quantitative effects on the results, leaving our overall
discussion and conclusions unaffected; furthermore, the 
corral's geometry can, in fact, be tuned to enhance/optimize 
the ME effects (depending on the parameters).



\begin{figure}[b]
\scalebox{0.72}{\includegraphics{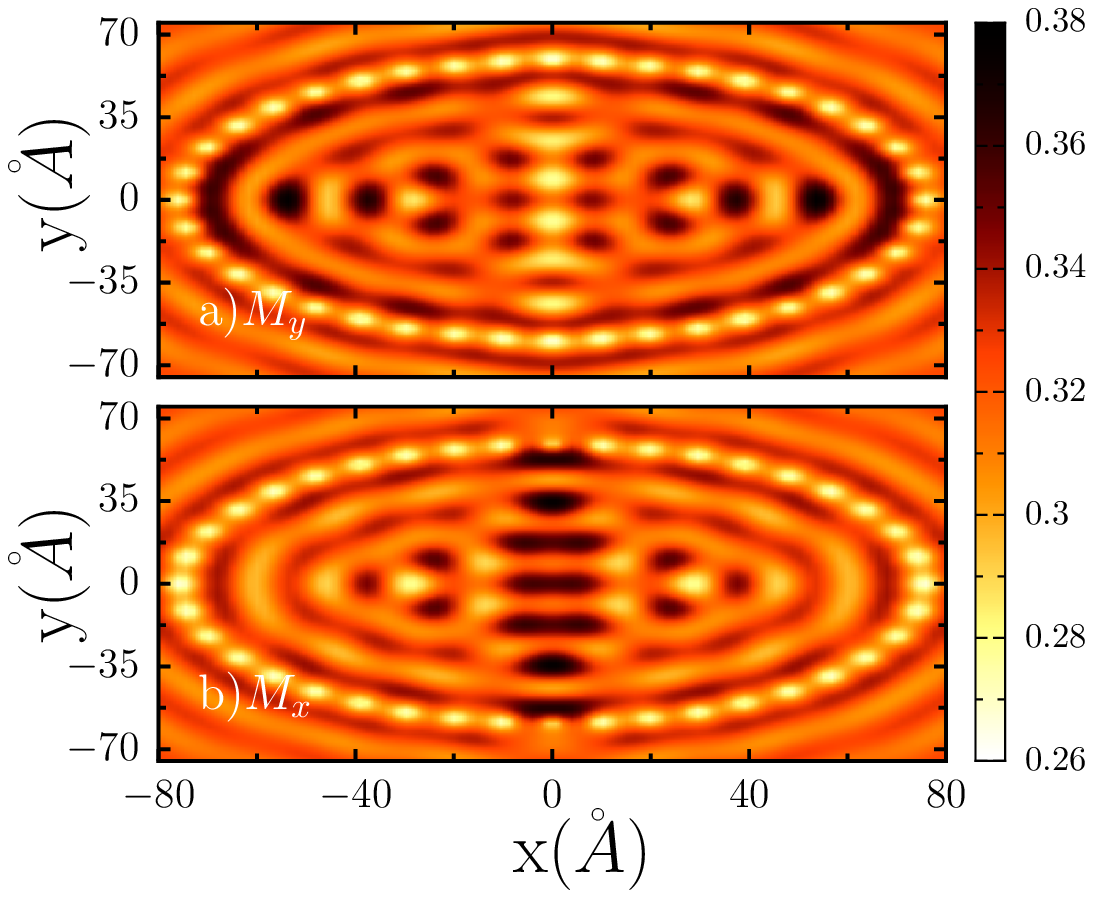} }
\vspace{-0.17in}
\caption{Spatial scan of the LDOS at $E_F$:
(a) ${\bf M}$=$|{\bf M}| \hat{y}$
(b) ${\bf M}$=$|{\bf M}| \hat{x}$.}
\label{fig:ellipsespace}
\end{figure}

\begin{figure}[t]
\scalebox{0.75}{\hspace{-0.31in} \includegraphics{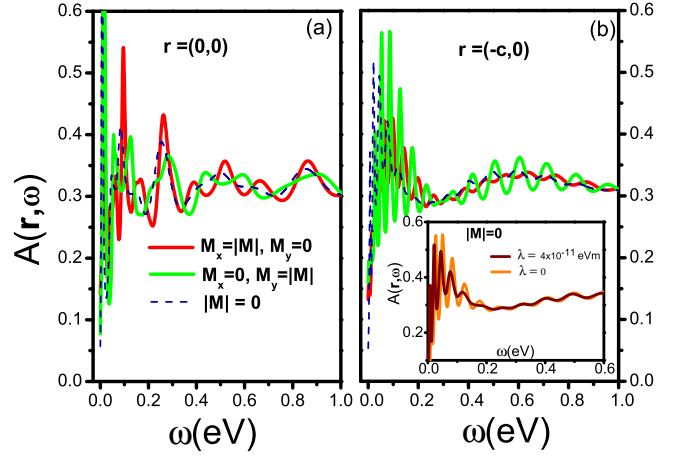}}
\vspace{-0.38in}
\caption{Energy dependence of the LDOS for different ${\bf M}$:
(a) ${\bf r}$=$(0,0)$ (b) ${\bf r}$=$(-c,0)$.
Inset: LDOS at $(-c,0)$ for $| {\bf M}|=0$ with and without SOC.}
\label{fig:ellipseenergy}
\end{figure}

We begin by discussing the electronic properties of the
QC (with $\hat{H}_1$=0).
Fig.~\ref{fig:ellipsespace} shows a spatial scan of the 
LDOS at $E_F$ for different directions of ${\bf M}$.
Notice how the LDOS changes as one changes the direction 
of ${\bf M}$ --- the magnetization of the QC's wall and 
the SOC give rise to strong ME effects at the atomic scale.
These effects are due to the breaking of SU(2) 
spin-rotation invariance by the SOC.  
Furthermore, the differences in the LDOS for 
${\bf M}$=$|{\bf M}| \hat{x}$ and 
${\bf M}$=$|{\bf M}| \hat{y}$ in
the elliptical QC are due to the breaking of rotational 
invariance --- for a circular QC, the DOS 
for ${\bf M}$=$|{\bf M}| \hat{x}$ 
and ${\bf M}$=$|{\bf M}| \hat{y}$ are identical. 
Fig.~\ref{fig:ellipseenergy} shows the energy dependence 
of the DOS at particular points in the QC --- changing 
the direction of ${\bf M}$ changes the energy dependence 
of the DOS; indeed, by carefully choosing the position in 
the QC, the changes can be quite pronounced  
[see Fig.~\ref{fig:ellipseenergy}(b)].
For reference, the DOS with $|{\bf M}|$=0 is also shown,
as well as the DOS with $\lambda$=0 (in the inset).

The ME effects exhibited by this system could find utility 
in a variety of applications that exploit the spatial and/or 
energy dependence of the QC's LDOS.
To illustrate the utility of the spatial dependence
of the LDOS, we place a spin-1/2 magnetic impurity at 
a position ${\bf R}_0$ inside the QC; we investigate 
how its low-energy properties depend on its position. 
We describe the magnetic impurity by the Hamiltonian
\begin{equation}
 \hat{H}_{1} = J_K~ \overline{\tau} \cdot {\bf S}({\bf R}_0) \ ,
\label{sdhammy}
\end{equation}
where $\overline{\tau}$ is the impurity's spin operator, 
${\bf S}({\bf R}_0)$ is the 2DEG's spin operator at the 
position ${\bf R}_0$, and $J_K$ is the exchange coupling 
between the impurity spin and the 2DEG.  In what follows,
we take $J_K$$>$$0$.\cite{SI}

\begin{figure}[b]
\scalebox{0.69}{\hspace{-0.27in} \includegraphics{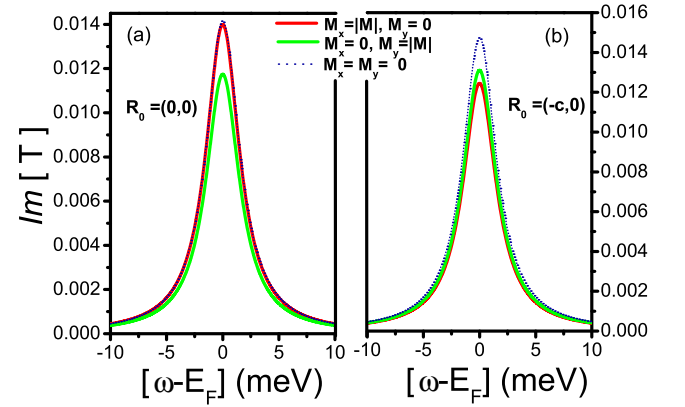}}
\vspace{-0.38in}
\caption{Imaginary part of the $T$-matrix due to a
magnetic impurity for different ${\bf M}$: 
(a) impurity at the origin ${\bf R}_0$=$(0,0)$
(b) impurity at the focus ${\bf R}_0$=$(-c,0)$.}
\label{fig:kondopeakellipse}
\end{figure}

Interestingly, this seemingly simple system exhibits 
nontrivial behavior in the infrared, due to quantum
fluctuations --- a strongly correlated state arises, where 
a cloud of conduction electrons forms a singlet with the 
impurity.\cite{hewson}
This strongly correlated state manifests itself via a
resonance at (or near) the Fermi energy, referred to as 
the Kondo resonance (KR).
This can be seen in Fig.~\ref{fig:kondopeakellipse}, where 
the imaginary part of the impurity's $T$-matrix is shown and, 
in particular, the KR appears.
Furthermore, the width of this KR represents the 
dynamically generated scale characteristic of this strongly 
correlated state, the Kondo temperature $T_K$.\cite{hewson}

\begin{figure}[t]
\scalebox{0.72}{\includegraphics{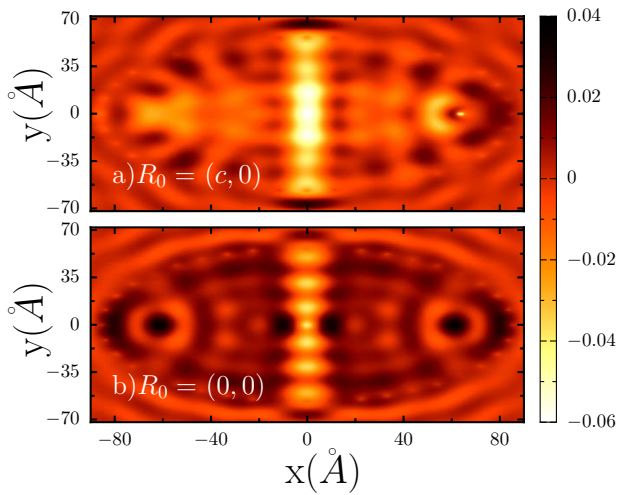}}
\vspace{-0.17in}
\caption{Spatial dependence of the LDOS difference at $E_F$
due to a magnetic impurity, 
$\delta A({\bf r}, E_F)$ (see text): (a) impurity at the 
focus ${\bf R}_0$=$(c,0)$ (b) impurity at the origin 
${\bf R}_0$=$(0,0)$.}
\label{fig:DOSkondo}
\end{figure}

The ME effects allow one to control the Kondo effect 
exhibited by the magnetic impurity in the QC; more 
generally, it allows one to control the magnetic 
properties of an atom placed in the QC.
Indeed, Fig.~\ref{fig:kondopeakellipse} shows that the KR
can be controlled by changing (the direction of) ${\bf M}$.
Furthermore, we see that the change in the KR depends on 
the position at which the impurity is placed\cite{uic} 
--- one can control the impurity's properties in a desired 
way by a judicious choice of its position.
It should also be noted that, besides impacting the 
properties of the magnetic impurity, the QC is also 
impacted by the magnetic impurity; 
the impurity's influence on the QC depends on its 
position.
This can be seen in Fig.~\ref{fig:DOSkondo}, where 
a spatial scan of the quantity 
$\delta A({\bf r},E_F)$=$A({\bf r},E_F)_{\hat{x}}$$-$$
A({\bf r},E_F)_{\hat{y}}$ --- the difference in the 
LDOS (at $E_F$) between  ${\bf M}$=$|{\bf M}| \hat{x}$ 
and ${\bf M}$=$|{\bf M}| \hat{y}$ --- is shown for 
different positions of the magnetic impurity.
[Results obtained for a circular QC are qualitatively 
similar to those obtained for an elliptical QC.]



\begin{figure}[t]
\scalebox{0.71}{
\hspace{-0.39in} \includegraphics{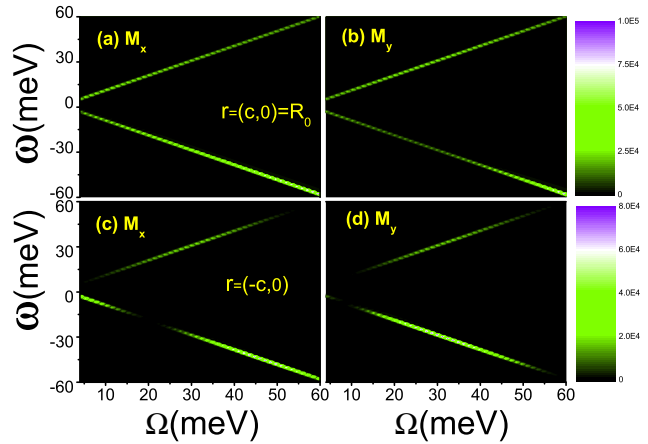}}
\vspace{-0.42in}
\caption{Maps of
$|\partial_\omega \delta A({\bf r};\omega)|$ 
due to the vibrational mode for 
${\bf M}$=$|{\bf M}| \hat{x}$ 
and ${\bf M}$=$|{\bf M}| \hat{y}$: 
(a) and (b) at ${\bf r}$=$(c,0)$=${\bf R}_0$; 
(c) and (d) at ${\bf r}$=$(-c,0)$, demonstrating the 
filtering of the mirage signal.}
\label{fig:vibration}
\end{figure}

As we saw above, the ME effects allow manipulation of both 
the spatial and energy dependence of the QC's LDOS.
As we now demonstrate, this change in the energy dependence 
of the DOS can be used for signal filtering.  
To this end, we place a molecule in the QC at the 
focus $(c,0)$, such that a vibrational mode (VM) of this 
molecule couples to the 2DEG of the QC; we investigate the 
image or "mirage" of the VM at the other focus $(-c,0)$.
We describe the VM by\cite{holstein}
\begin{equation}
 \hat{H}_{1} = \frac{1}{2} p^2 
   + \frac{1}{2} \Omega^2 x^2 + g~x~ n({\bf R}_0) \ ,
\end{equation}
where $x$ ($p$) is the position (momentum) operator of
the VM, $\Omega$ is its characteristic frequency, 
$n({\bf R}_0)$ is the 2DEG's density operator at the
position ${\bf R}_0$=$(c,0)$, 
and $g$ describes the coupling between the VM and the 2DEG.
In what follows, we assume the coupling between the 2DEG 
and the VM to be weak.\cite{SI}

The QC enables signals to be transmitted between 
foci;\cite{corralexp} it also provides a 
"cloak of invisibility", similar to what has been achieved 
with electromagnetic fields,\cite{metamaterials} where 
objects were made invisible within a certain frequency band.
More specifically, the electronic properties of the QC 
govern the energy regime in which a signal transmitted
from one focus can be observed at the other focus;
in particular, signals within a certain frequency range
can be hidden from observation.\cite{manobalat}
This cloaking can be seen in Fig.~\ref{fig:vibration}, 
where a density plot of the energy derivative of the LDOS 
$|\partial_\omega A({\bf r};\omega)|$ is shown --- even 
when the VM is visible at the focus in which it is sitting 
(Fig.~\ref{fig:vibration}(a) and (b)), the QC cloaks it 
from observation at the other focus for $\Omega$ within 
a certain range 
(Fig.~\ref{fig:vibration}(c) and (d)).\cite{manobalat}
In this system, the ME effects allow the range over which 
cloaking occurs to be controlled by changing the 
orientation of ${\bf M}$ --- compare the results for 
${\bf M}$=$|{\bf M}| \hat{x}$ and ${\bf M}$=$|{\bf M}|\hat{y}$.  
Said in another way, the ME effects allow one to filter the 
signal transmitted from one focus to the other.



\begin{figure}[t]
\scalebox{0.65}{\hspace{-0.41in} \includegraphics{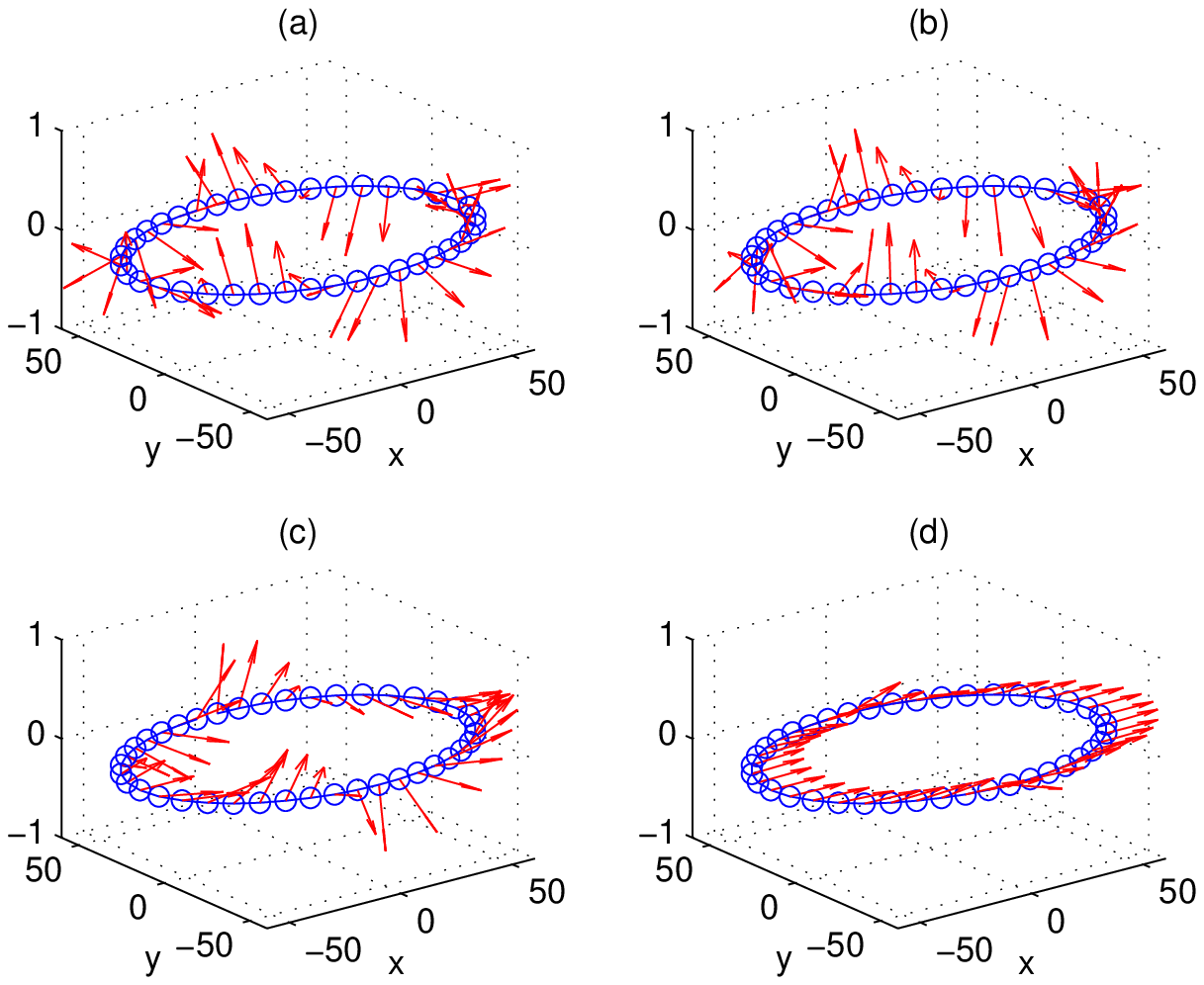}}
\vspace{-0.33in}
\caption{Evolution of the magnetization with the
external magnetic field ${\bf h}$: (a) $|{\bf h}|$=0
(b) $|{\bf h}|$=0.04$E_0$ (c) $|{\bf h}|$=0.12$E_0$
(d) $|{\bf h}|$=0.24$E_0$ where $E_0$=$\pi(\rho_0 J)^2$$E_F$.}
\label{fig:magnetization}
\end{figure}

Up to now, we have considered the system's electronic
properties, assuming the QC's wall to be FM ordered.  
We have also considered the magnetic properties of the 
wall and, in particular, the wall's magnetic ordering
tendencies;\cite{SI}
as the QC's wall is one-dimensional, fluctuations will 
suppress ordering, and an external field ${\bf h}$ is 
necessary to stabilize the order. 
Fig.~\ref{fig:magnetization} shows results for the QC's 
magnetization for various values of ${\bf h}$.
We see that the moments are disordered at zero field and, hence, 
the magnetization is zero; a nonzero magnetization is obtained
as ${\bf h}$ is increased.
We have found that a FM aligned wall is obtained from readily
accessible magnetic fields --- 
$|{\bf h}|$$\simeq$$0.2 \pi (\rho_0 J)^2 E_F$; for reasonable 
values of the parameters, this gives 
$|{\bf h}|$=${\cal O}$(1meV).
[It is worth noting such values are considerably lower than 
Kondo temperatures that have been observed from atoms/molecules 
on surfaces.\cite{kondosurface}]
Furthermore, we have found the geometry can, in fact, be 
optimized, so that ferromagnetic ordering occurs at extremely 
small values of ${\bf h}$.



This work demonstrates proof of principle of the functionality 
afforded by ME effects at the atomic scale; indeed, we were 
able to control the properties of systems placed inside the QC 
as well as their electronic signals/signatures.
With a FM aligned wall, the ME effects allowed us to control 
the magnetic properties of atoms placed inside the QC, as well 
as to filter transmitted signals; 
different magnetization patterns for the 
wall,\cite{surfacemagnetchiral} as well as different QC 
geometries, could provide further flexibility and control. 
By placing several atoms/molecules inside the QC, one could 
engineer devices where the ME effects allow the 
entanglement\cite{entanglereview} between atoms/molecules to 
be manipulated.  
It should also be mentioned that the ME effects allow 
manipulation of the properties of the QC's wall
--- the QC's wall itself provides a unique magnetic system
with interesting properties, which could also afford  
means of transmitting and manipulating 
information.\cite{surfacechain,surfaceRKKY}



EHK acknowledges the warm hospitality of the
Instituto de F\'isica T\'eorica (Madrid, Spain), where 
most of this work was performed.  SEU acknowledges 
support from AvH Stiftung, and the hospitality of the 
Dahlem Center for Complex Quantum Systems at FU-Berlin.
This work was supported by 
NSF MWN/CIAM and PIRE grants (ATN and SEU),
the Spanish Grants TOQATA and QUAGATUA (JRL), 
and the Spanish Grant FIS2009-11654 (JRL and EHK).




\newpage

\begin{widetext}

\begin{center}
\large
{\bf Quantum manipulation via atomic-scale 
magnetoelectric effects: \\
Supplementary Information}
\end{center}


\section{The System and Hamiltonian}

We consider a quantum corral made of magnetic atoms on 
a metallic surface with spin-orbit coupling (SOC).  
The Hamiltonian is $\hat{H}_{\rm QC}$=$\hat{H}_0$+$\hat{V}$,
where $\hat{H}_0$ describes the two-dimensional electron
gas (2DEG) of the surface, and
$\hat{V}$ describes the coupling of the 2DEG to the
magnetic atoms.
In second quantized form, the Hamiltonian for the 2DEG is
\begin{equation}
 \hat{H}_0 = \int d{\bf r}~ \psi^{\dagger}({\bf r})
  \left[ \frac{1}{2m} {\bf p}^2 + \lambda~ \hat{z}
  \cdot \left( {\bf p} \times \overline{\sigma} \right)
  \right] \psi({\bf r}) \ ,
\label{2DEGhamsupp}
\end{equation}
where $\psi^{\dagger}({\bf r})$ is a two-component field
operator for the 2DEG $\psi^{\dagger}({\bf r}) =
(\psi^{\dagger}_{\uparrow}({\bf r}),
\psi^{\dagger}_{\downarrow}({\bf r}))$,
and $\{ \sigma^{\mu} \}$ are the Pauli matrices;
the coupling of the 2DEG to the magnetic atoms is
\begin{equation}
 \hat{V} = \sum_i \psi^{\dagger}({\bf r}_i)
  \left( V_0 + \frac{J}{2} {\overline \tau}_i 
  \cdot {\overline \sigma} \right) \psi({\bf r}_i)  \ ,
\label{couplingsupp}
\end{equation}
where $\overline{\tau}_i$ is the spin operator for the 
magnetic moment of the $i^{\rm th}$ atom, $V_0$ describes 
the potential scattering, and $J$ is the exchange coupling 
between the 2DEG and the magnetic atoms.
[As before, $\{ \sigma^{\mu} \}$ are the Pauli matrices.]

As we are interested in the case where the wall is 
ferromagnetically ordered, we treat the magnetic
moments of the atoms as classical variables; then 
$(J/2) \langle \overline{\tau}_i 
\rangle$$\rightarrow$${\bf M}$.
The quantity entering in the calculations is 
$(\pi \rho_0)$$|{\bf M}|$.  To estimate this quantity,
we consider, for concreteness, a spin-5/2 moment (which 
is relevant to e.g. ${\rm Mn}^{2+}$ atoms); we consider 
the physically reasonable value $\rho_0 J$=$0.2$ --- we 
obtain $(\pi \rho_0)$$|{\bf M}|$$\simeq$0.785.
Motivated by this value, in our calculations we used
$(\pi \rho_0)$$|{\bf M}|$=0.5.
As noted in the text, however, our results are robust,
as the corral's geometry can be tuned to enhance/optimize
the magnetoelectric effects.


\section{Scattering Formalism}

The QC's GF can be written as 
\begin{equation}
 G({\bf r},{\bf r}'; \omega) 
   = G_0({\bf r},{\bf r}'; \omega)
   + G_0({\bf r},{\bf R}_0; \omega) \hat{T}(\omega)
     G_0({\bf R}_0,{\bf r}'; \omega)  \ .
\label{fullgreenssupp}
\end{equation}
In Eq.~\ref{fullgreenssupp}, the $T$-matrix $\hat{T}(\omega)$ 
describes the influence of $\hat{H}_1$.  Furthermore, 
$G_0({\bf r},{\bf r}'; \omega)$ is the bare GF of the QC 
--- it is the GF in the absence of $\hat{H}_1$; it is 
determined by the Dyson equation
\begin{eqnarray}
  & &  G_0({\bf r}, {\bf r}'; \omega) 
  = G_{00}({\bf r}, {\bf r}'; \omega) 
  \label{finaldyson}  \\  & & \hspace{0.25in} 
  + \sum_i G_{00}({\bf r}, {\bf r}_i; \omega) 
        \left( V_0~ I + {\bf M} \cdot {\overline \sigma} \right) 
         G_0({\bf r}_i, {\bf r}'; \omega)
\nonumber
\end{eqnarray}
with $G_{00}({\bf r}, {\bf r}'; \omega)$ being the free-particle
GF i.e. the GF in the absence of the QC's wall (and also $\hat{H}_1$).
$G_{00}({\bf r}, {\bf r}'; \omega)$ is given by\cite{SOgreens1supp,
SOgreens2supp} (for ${\bf r}$$\neq$${\bf r}'$)
\begin{equation}
 G_{00}({\bf r},{\bf r}';\omega) = G^{00}_0(R;\omega) ~I 
 + G^{00}_1(R;\omega) \left( \begin{array}{c c}
    0 & -i \exp(-i\theta) \\
    i \exp(i\theta) & 0 \end{array} \right)
\end{equation}
where
\begin{subequations}
\begin{eqnarray}
 G^{00}_{0}(R;\omega) & = & -i \frac{m}{4} \left\{ 
    \left( 1 + \frac{\lambda m}{k} \right) H_0[R (k+\lambda m)]  
  + \left( 1 - \frac{\lambda m}{k} \right) H_0[R (k-\lambda m)] 
 \right\}  \label{freediagsupp}  \\
 G^{00}_{1}(R;\omega) & = & - \frac{m}{4} \left\{ 
    \left( 1 + \frac{\lambda m}{k} \right) H_1[R (k+\lambda m)]  
  - \left( 1 - \frac{\lambda m}{k} \right) H_1[R (k-\lambda m)]
 \right\}   \label{freespinflipsupp} 
\end{eqnarray}
\end{subequations}
with $H_0(x)$=$J_0(x)$+$i N_0(x)$, $H_1(x)$=$J_1(x)$+$i N_1(x)$,
and $\exp(i \theta)$=$[(x - x') + i (y - y')]/|{\bf r} - {\bf r}'|$.
In the above equations,
$J_0(x)$ and $J_1(x)$ ($N_0(x)$ and $N_1(x)$) are the Bessel 
function (Neumann function) of order-zero and order-1, 
respectively.\cite{gradshteyn}
Furthermore, $R = |{\bf r} - {\bf r}'|$ and
$k$ is such that $k^2/2m = \omega + (\lambda m)^2/2m$.


\section{Magnetic Impurity}

We consider a spin-1/2 magnetic impurity placed in the QC;
the Hamiltonian is
\begin{equation}
 \hat{H}_{1} = J_K~ {\overline \tau} \cdot 
 \psi^{\dagger}({\bf R}_0) 
 \left( {\overline \sigma}/2 \right) \psi({\bf R}_0)
\label{sdhammysupp}
\end{equation}
where $\overline{\tau}$ is the impurity's spin operator.  
[$J_K$$>$0.]
To proceed, we treat ${\overline \tau}$ with a 
fermion representation --- we write
$\overline{\tau}$=$(1/2) f^{\dagger}
 \overline{\sigma} f^{\phantom \dagger}$
where $f^{\dagger}$ is the two-component spinor
$f^{\dagger}$=$
(f^{\dagger}_{\uparrow},f^{\dagger}_{\downarrow})$;
the $f$-fermions satisfy the constraint 
$f^{\dagger}$$f$=1.
Then, Eq.~\ref{sdhammysupp} can be written as
\begin{equation}
 \hat{H}_1 = -\frac{J_K}{2} 
 \left( \psi^{\dagger}({\bf R}_0)  
        f^{\phantom \dagger}\hspace{-0.1cm}\right) 
 \left( f^{\dagger} \psi({\bf R}_0) \right) \ .
\label{coqblinsupp}
\end{equation}

As we are interested in the infrared fixed point of 
Eq.~\ref{sdhammysupp} (or, equivalently, Eq.~\ref{coqblinsupp}), 
we employ mean-field theory\cite{hewsonsupp} --- the infrared 
properties are determined by the effective Hamiltonian
\begin{equation}
 \hat{H}_{\rm eff} = \lambda~ f^{\dagger} f 
  + \chi~ \psi^{\dagger}({\bf R}_0) f
  + \chi~ f^{\dagger}\psi({\bf R}_0) \ ,
\label{HShammysupp}
\end{equation}
where $\chi$ and $\lambda$ are (constants) determined
self-consistently via
\begin{equation}
 \frac{4}{J_K} \chi = - \left\langle
    \psi^{\dagger}({\bf R}_0) f
  + f^{\dagger} \psi({\bf R}_0)  \right\rangle
  \ \ \ , \ \ \   
  \left\langle f^{\dagger} f \right\rangle = 1 
   \ \ \ .
\label{meanfieldsupp}
\end{equation}
Then, the $T$-matrix in Eq.~\ref{fullgreenssupp} is proportional
to the $f$-fermions' retarded GF:\cite{hewsonsupp}
$\hat{T}(\omega) = \chi^2~ G^f(\omega)$, where
\begin{equation}
 G^f(\omega) = -i \Theta(t) \left\langle
 \{ f(t),f^{\dagger} \} \right\rangle(\omega)  \ .
\end{equation}
 

\section{Vibrational Mode}

\begin{figure}[ht]
\scalebox{.32}{\includegraphics{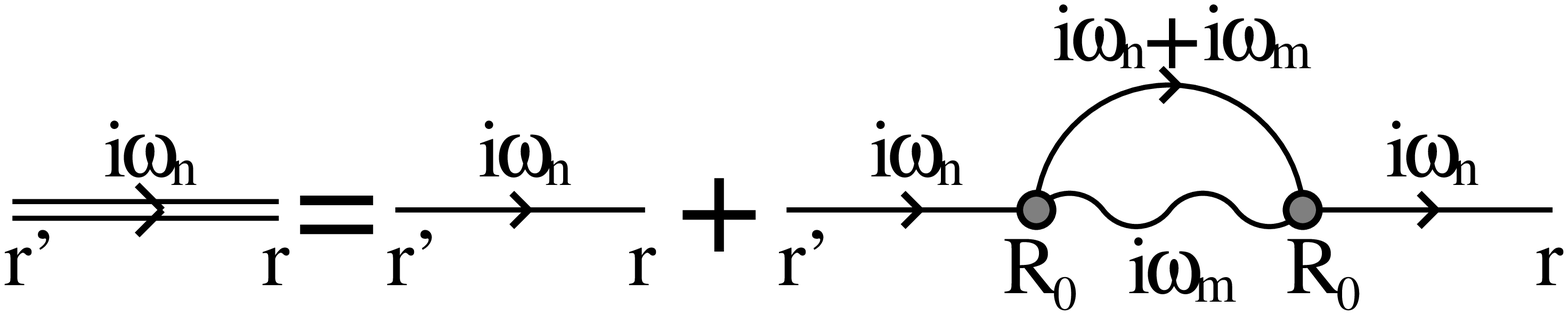} }
\caption{QC's GF, taking into account the VM.
The single solid (wavy) line denotes the QC's 
(VM's) bare GF.}
\label{fig:diagram}
\end{figure}

Assuming the coupling between the VM and the 2DEG
to be weak (i.e. $g$ is small), we take the VM into 
account perturbatively.
To lowest nontrivial order, the QC's GF
(taking into account the VM) is given by the diagram in 
Fig.~\ref{fig:diagram}.
Performing the Matsubara sum and analytically continuing
to real frequencies, the $T$-matrix (in 
Eq.~\ref{fullgreenssupp}) is given by 
\begin{equation}
 \hat{T}(\omega) = g^2 \int \frac{d\nu}{2\pi} A_0(\nu)
  \left\{ \frac{[1 + n(\Omega) - f(\nu)]}
     {\omega - \nu - \Omega + i\delta} 
   + \frac{[n(\Omega) + f(\nu)]}
     {\omega - \nu + \Omega + i\delta} \right\}  \ \ ,
\label{spectralsupp}
\end{equation} 
where $A_0(\nu)$ is the corral's spectral function 
at the site ${\bf R}_0$ ---
$A_0(\nu)$ is such that the Matsubara GF can be 
written as\cite{mahansupp}
\begin{equation}
 G_0({\bf R}_0, {\bf R}_0; i\omega_n) = \int \frac{d\nu}{2\pi}~
  \frac{A_0(\nu)}{i\omega_n - \nu} \ \ .
\nonumber 
\end{equation} 


\section{Magnetic Properties of the Quantum Corral's Wall}

As we are interested in the magnetic properties of the 
QC's wall, we return to Eqs.~\ref{2DEGhamsupp} and 
\ref{couplingsupp} --- we integrate out the 2DEG and 
obtain an effective spin Hamiltonian;\cite{spinglassbooksupp} 
the leading interaction generated\cite{RGsupp} is given 
by\cite{SOgreens1supp,rkky2Dsupp}
\begin{equation}
 \hat{H}_{\rm spin} = -\frac{J^2}{4\pi} \sum_{i<j}
 \int d\omega~ f(\omega)~ {\rm Im} \left\{~{\rm Tr} \left[ 
   \left(\overline{\tau}_i \cdot \overline{\sigma}\right) 
   G_{0}({\bf r}_i,{\bf r}_j; \omega)
   \left(\overline{\tau}_j \cdot \overline{\sigma}\right)
   G_{0}({\bf r}_j,{\bf r}_i; \omega)
 \right]~ \right\}  \ ,
\label{RKKYintegralsupp}
\end{equation} 
where $f(\omega)$ is the Fermi function, and 
$G_{0}({\bf r}_i,{\bf r}_j; \omega)$ is the retarded GF 
of the 2DEG in the absence of the exchange coupling $J$.
To proceed efficiently, we employ a commonly used 
approximation, namely approximating 
$G_{0}({\bf r}_i,{\bf r}_j; \omega)$ in 
Eq.~\ref{RKKYintegralsupp} by 
$G_{00}({\bf r}_i,{\bf r}_j; \omega)$.\cite{spinglassbooksupp,
freeapproximationsupp}
We have checked that {\sl along the corral's wall}, 
$G_{0}({\bf r}_i,{\bf r}_j; \omega)$ is similar to 
$G_{00}({\bf r}_i,{\bf r}_j; \omega)$.
[Of course this approximation fails inside the corral.]
We obtain
\begin{equation}
 \hat{H}_{\rm spin} = \sum_{i<j} 
    K_{ij}~ \overline{\tau}_i \cdot \overline{\tau}_j
 +  D_{ij}~ {\bf Q}_{ij} \cdot \left( 
        \overline{\tau}_i \times \overline{\tau}_j \right)
 + J^0_{ij} ~ \left( {\bf Q}_{ij} \cdot \overline{\tau}_i \right) 
              \left( {\bf Q}_{ij} \cdot \overline{\tau}_j \right)  \ ,
\label{RKKYfreeelectronsupp}
\end{equation}  
where 
${\bf Q}_{ij}$=$\hat{z} \times \hat{R}_{ij}$ with
$\hat{R}_{ij}$=$({\bf R}_i - {\bf R}_j)/|{\bf R}_i - {\bf R}_j|$
($\hat{z}$ is the unit vector perpendicular to the 2DEG),
and the couplings are given by
\begin{subequations}
\begin{eqnarray}
 K_{ij} & = & -\frac{J^2}{2\pi} \int_0^{E_F} 
  \hspace{-0.13in} d\omega~   
  {\rm Im} \left\{ \left[ 
    G^{00}_{0}(|{\bf r}_i - {\bf r}_j|;\omega) \right]^2
  + \left[ G^{00}_{1}(|{\bf r}_i - {\bf r}_j|;\omega) \right]^2 
  \right\} \ ,  \nonumber \\
 D_{ij} & = & \frac{J^2}{\pi} \int_0^{E_F} 
  \hspace{-0.13in} d\omega~
  {\rm Re} \left\{ G^{00}_{0}(|{\bf r}_i - {\bf r}_j|;\omega) 
    G^{00}_{1}(|{\bf r}_i - {\bf r}_j|;\omega)  \right\}  \ ,   
   \nonumber \\
 J^0_{ij} & = & \frac{J^2}{\pi} \int_0^{E_F} 
  \hspace{-0.13in} d\omega~
  {\rm Im} \left\{ 
    \left[ G^{00}_{1}(|{\bf r}_i - {\bf r}_j|;\omega) \right]^2
    \right\}  \ ,  \nonumber
\end{eqnarray}
\end{subequations}
where $E_F$ is the Fermi energy.
[$G^{00}_{0}(|{\bf r}_i - {\bf r}_j|;\omega)$ and
$G^{00}_{1}(|{\bf r}_i - {\bf r}_j|;\omega)$ are given in
Eqs.~\ref{freediagsupp} and \ref{freespinflipsupp}.]

The magnetic properties of the QC's wall were determined
by Monte Carlo simulations of Eq.~\ref{RKKYfreeelectronsupp} 
--- the magnetic moments were treated as classical 
three-dimensional vectors of unit length; the minimum energy 
state of Eq.~\ref{RKKYfreeelectronsupp} was obtained via a 
simulated annealing procedure. 


\end{widetext}



\begin{thebibliography}{100}
\bibitem{RGsupp}{This is the most relevant interaction 
generated in the renormalization group sense.}
\end{thebibliography}
\end{document}